\documentclass[a4paper,fleqn,usenatbib]{mnras}
\pdfoutput=1
\usepackage{mathptmx}

\usepackage[T1]{fontenc}
\usepackage{ae,aecompl}

\usepackage{graphicx}	
\usepackage{amsmath}	
\usepackage{amssymb}	
\usepackage{hyperref}
\usepackage{tabularx}

\newcommand{\kms}{\ensuremath{\mathrm{km}\,\mathrm{s}^{-1}}}

\newcommand{\Vd}{\ensuremath{V_{2.2}}}
\newcommand{\Ve}{\ensuremath{V_{\rm 2 R_e}}}
\newcommand{\Vm}{\ensuremath{V_{\rm max}}}
\newcommand{\Vf}{\ensuremath{V_{\rm f}}}
\newcommand{\Wp}{\ensuremath{W_{\rm P20}}}
\newcommand{\Wm}{\ensuremath{W_{\rm M50}}}
\newcommand{\Wmc}{\ensuremath{W_{\rm M50c}}}

\newcommand{\hi} {{\rm H\,{\footnotesize\rm I}}}

\title[The baryonic Tully-Fisher relation for different velocity definitions]{The baryonic Tully-Fisher relation for different velocity definitions and implications for galaxy angular momentum}

\author[F. Lelli et al.]{Federico Lelli$^1$\thanks{ESO Fellow. E-mail: flelli@eso.org}, Stacy S. McGaugh$^2$, James M. Schombert$^3$, 
Harry Desmond$^4$, \newauthor Harley Katz$^4$\\
$^1$European Southern Observatory, Karl-Schwarschild-Strasse 2, 85748 Garching bei M\"{u}nchen, Germany\\
$^2$Department of Astronomy, Case Western Reserve University, Cleveland, OH 44106, USA\\
$^3$Department of Physics, University of Oregon, Eugene, OR 97403, USA\\
$^4$Astrophysics, University of Oxford, Denys Wilkinson Building, Keble Road, Oxford OX1 3RH, UK
}

\date{Accepted XXX. Received YYY; in original form ZZZ}

\pubyear{201}

\begin{document}
\label{firstpage}
\pagerange{\pageref{firstpage}--\pageref{lastpage}}
\maketitle

\begin{abstract}
We study the baryonic Tully-Fisher relation (BTFR) at $z\simeq0$ using 153 galaxies from the SPARC sample. We consider different definitions of the characteristic velocity from \hi\ and H$\alpha$ rotation curves, as well as \hi\ line-widths from single-dish observations. We reach the following results: (1)\,The tightest BTFR is given by the mean velocity along the flat part of the rotation curve. The orthogonal intrinsic scatter is extremely small ($\sim$6$\%$) and the best-fit slope is $3.85\pm0.09$, but systematic uncertainties may drive the slope from 3.5 to 4.0. Other velocity definitions lead to BTFRs with systematically higher scatters and shallower slopes. (2)\,We provide statistical relations to infer the flat rotation velocity from \hi\ line-widths or less extended rotation curves (like H$\alpha$ and CO data). These can be useful to study the BTFR from large \hi\ surveys or the BTFR at high redshifts. (3)\,The BTFR is more fundamental than the relation between angular momentum and galaxy mass (the Fall relation). The Fall relation has about 7 times more scatter than the BTFR, which is merely driven by the scatter in the mass-size relation of galaxies. The BTFR is already the ``fundamental plane'' of galaxy discs: no value is added with a radial variable as a third parameter.
\end{abstract}

\begin{keywords}
dark matter -- galaxies: kinematics and dynamics -- galaxies: spiral --- galaxies: dwarf
\end{keywords}



\section{Introduction}

In a seminal paper, \citet{Tully1977} discovered a tight correlation between the luminosity of a galaxy and the width of its global \hi\ profile. This Tully-Fisher (TF) relation has been widely used to determine galaxy distances, estimate the value of the Hubble constant, and study local galaxy flows \citep[e.g.,][]{Sandage1976, Tully1977, Tully1988, Sorce2013}. Subsequently, the relation has also become a major tool to test galaxy formation models in a $\Lambda$ Cold Dark Matter ($\Lambda$CDM) cosmology \citep[e.g.,][]{Mo1998, Desmond2015}, as well as alternative theories of modified gravity and/or modified inertia \citep[e.g.,][]{Milgrom1983, Sanders1990, McGaugh2012}.

Both luminosity and \hi\ line-width are proxies for more fundamental quantities. The luminosity relates to the stellar mass of the galaxy, while the \hi\ line-width is an indicator of the circular velocity. The stellar-mass TF relation, however, breaks down for velocities smaller than $\sim$100 \kms \citep{McGaugh2000}. This regime is dominated by gas-rich dwarf galaxies, in which the cold gas mass ($M_{\rm g}$) is comparable to or even larger than the stellar mass ($M_\star$). When the stellar mass is replaced with the total baryonic mass ($M_{\rm b} = M_\star + M_{\rm g}$), one recovers a single linear relation over $\sim$5 decades in $M_{\rm bar}$ \citep{McGaugh2000, McGaugh2005, Lelli2016a, Iorio2017}.

Nowadays there is little doubt that the baryonic mass provides the most fundamental form of this relation: the baryonic TF relation (BTFR). Some confusion, however, persists regarding the definition of the circular velocity. Different authors often use different velocity definitions depending on the available data. This leads to BTFRs with different slopes, intercepts, and scatters, which complicate the comparison with cosmological simulations of galaxy formation \citep[e.g.,][]{Brook2016}. Moreover, this affects the comparison between TF relations at $z\simeq0$ with those at higher $z$, leading to contradictory conclusions about its possible cosmic evolution \citep[see discussion in][]{Turner2017}.

The radial extent of the kinematical tracer plays a crucial role. Ionized gas (like H$\alpha$) and molecular gas (like CO) generally probe circular velocities within the bright stellar disc, which do not always reach the flat portion of the galaxy rotation curve. Atomic gas (\hi), instead, probes circular velocities out to large radii since \hi\ discs are generally more extended than stellar discs. On the other hand, while the kinematics of ionized and molecular gas can be studied up to $z\simeq3-4$, \hi\ observations are currently limited to $z\simeq0$.

The kinematical status of the velocity tracer is also important. In general, atomic and molecular gas display low velocity dispersions ($\sigma \lesssim 10$ \kms) and probe kinematically-cold discs with $V_{\rm rot}/\sigma \simeq 5-30$. Thus, the observed rotation velocity $V_{\rm rot}$ directly traces the circular velocity $V_{\rm circ}$ of a test particle in the galaxy potential. Corrections for pressure support (asymmetric drift) are important only for the smallest dwarf galaxies with $V_{\rm rot}\simeq20$ \kms \citep[e.g.,][]{Lelli2012}. Conversely, ionized gas can display larger velocity dispersions and is more strongly affected by stellar feedback than neutral gas, so corrections for pressure support can be important. Some authors, indeed, advocate to use of the kinematical quantity $S_{0.5} = \sqrt{0.5 V_{\rm rot}^2 + \sigma^2}$ in H$\alpha$ surveys in order to locate ``pressure-dominated'' and ``rotation-dominated'' galaxies on the same plane \citep{Cortese2014}. In practice, $S_{0.5}$ is a rough asymmetric-drift correction, where the contributions from anisotropy, velocity dispersion profile, and density profile are encoded into a simple constant (cf. with Eq.\,A.6 in \citealt{Lelli2014}).

Finally, there are key differences between spatially resolved data, like interferometric \hi\ observations, and spatially unresolved data, like single-dish \hi\ observations. The latter provide the global \hi\ profile, which is a spatial convolution between the galaxy rotation curve and the \hi\ density profile. The width of the global \hi\ profile is roughly twice the circular velocity, but its value can be skewed towards the velocities of the high-density gas in the inner galaxy regions. Indeed, velocities from spatially integrated \hi\ profiles can be similar to H$\alpha$ velocities \citep{Courteau1997} but different from the outer \hi\ velocities from spatially resolved rotation curves \citep{Verheijen2001b}, leading to systematic differences.

Up-coming facilities will boost the study of the BTFR. At $z\simeq0$, radio interferometers with phased-array feeds, such as APERTIF and ASKAP, are expected to provide \hi\ rotation curves for thousands of galaxies and \hi\ line-widths for hundreds of thousands \citep{Duffy2012, Adams2018}. At high $z$, the James Webb Space Telescope will provide accurate stellar masses from near-infrared photometry as well as rotation curves from integral-field spectroscopy of ionised gas, exceeding the current limits of ground-based observations. It is crucial, therefore, to quantify the systematic effects that different velocity measurements (such as \hi\ line-widths versus rotation curves) may have on the BTFR slope, normalization, and scatter.

In this paper, we investigate these issues using the Spitzer Photometry and Accurate Rotation Curve (SPARC) database \citep{Lelli2016b}. We consider different velocity definitions from spatially-resolved rotation curves as well as \hi\ line-widths from single-dish observations. This extends previous studies \citep{Verheijen2001b, Noordermeer2007b, Ponomareva2017} with a larger galaxy sample that covers a broader dynamic range in baryonic mass. The SPARC galaxy sample and the data analysis are described in Sect.\,\ref{sec:Data}. The properties of the various BTFRs are described in Sect.\,\ref{sec:BTFR}. Statistical corrections between different velocity measurements are given in Sect.\,\ref{sec:sys}. The BTFR residuals and their relation to the galaxy angular momentum are discussed in Sect.\,\ref{sec:residuals}. Our results are summarized in Sect.\,\ref{sec:Conc}.

\section{Data Analysis}\label{sec:Data}

The SPARC database provides \hi\ rotation curves accumulated over three decades of radio interferometry (see references in \citealt{Lelli2016b}) and homogeneous near-infrared photometry from $Spitzer$ [3.6] images \citep[e.g.,][]{Schombert2014b}. To date this is the largest sample of galaxies with both extended \hi\ rotation curves and [3.6] photometry. For $\sim$30$\%$ of the sample, we also have high-resolution H$\alpha$ rotation curves, which are combined with the \hi\ data to better sample the inner galaxy regions. All data are publicly available at \href{http://astroweb.cwru.edu/SPARC/}{astroweb.cwru.edu/SPARC/}.

The SPARC sample is representative for galaxies in the field, in nearby groups, and in diffuse clusters (such as Ursa Major). Galaxies in dense X-ray emitting clusters (such as Virgo) are not represented in SPARC because extended \hi\ rotation curves are not available, as these galaxies tend to be gas poor or to have truncated \hi\ discs \citep{Chung2009}. This does not introduce any significant bias since only a few percent of galaxies live in clusters and there is no evidence that the BTFR varies with environment \citep{Mocz2012}.

The total sample comprises 175 galaxies, but we use a sub-sample of 153 objects as we did in several publications \citep{McGaugh2016, Lelli2017}. This excludes 22 objects with low inclinations ($i < 30^{\circ}$), where the geometric corrections to the rotation velocities are large and uncertain, as well as 7 objects with low quality rotation curves (quality flag $Q=3$), which manifestly do not trace the equilibrium gravitational potential of the galaxy. These basic quality criteria do not introduce any significant bias.

\subsection{Velocity Definitions}\label{sec:VelDef}

In this paper we consider several velocity definitions that are often used in the literature:
\begin{enumerate}
 \item The line-width \Wp\ measured at 20$\%$ of the peak flux density of the global \hi\ profile. This is the classic velocity definition employed by \citet{Tully1977}. \Wp/2 roughly corresponds to the circular velocity in the outer parts of the \hi\ disc (after inclination correction).
 \item The line-width \Wm\ measured at 50$\%$ of the mean flux density of the global \hi\ profile \citep{Springob2005}. In principle, this definition depends less strongly on the intrinsic \hi\ density distribution than \Wp\ since it considers the mean flux density rather than the peak. However it depends on the specification of the velocity window for the summation of the flux. \citet{Courtois2009} find that $\Wm<\Wp$ with a mean difference of $\sim$10 \kms. This suggests that $\Wm$ probes velocities at slightly smaller radii than $\Wp$.
\item The circular velocity \Vd\ measured at $2.2\,R_{\rm d}$, where $R_{\rm d}$ is the disc scale length from an exponential fit to the outer luminosity profile. The motivation beyond this definition is that a self-gravitating disc with a purely exponential profile reaches the maximum circular velocity at 2.2\,$R_{\rm d}$ \citep{Freeman1970}. This definition is often used by H$\alpha$ observers \citep[e.g.,][]{Courteau1997, Courteau1999, Reyes2011}.
\item The circular velocity \Ve\ measured at $2\,R_{\rm e}$, where $R_{\rm e}$ is the effective radius encompassing half of the galaxy luminosity (sometimes also indicated as $R_{50}$ or $R_{\rm h}$). This radius is measured in a non-parametric way: there is no need to fit the luminosity profile with a prescribed function. This definition is advocated by \citet{Romanowsky2012} to estimate the stellar angular momentum of both late-type and early-type galaxies. For a pure exponential disc $2\,R_{\rm e} \simeq 3.2 R_{\rm d} \simeq R_{80}$, where $R_{80}$ is the radius encompassing 80$\%$ of the galaxy luminosity  \citep{Pizagno2007, Reyes2011}. \Ve\ permits a comparison to BTFR studies at high-$z$, which usually measure velocities at $\sim3\,R_{\rm d}$ or $R_{80}$ \citep[e.g.,][]{Turner2017}.
 \item The circular velocity $V_{\max}$ measured at the peak of the observed rotation curve. Since galaxy discs can show significant deviations from exponential profiles, it often occurs that $V_{\rm max} > V_{2.2}$. This definition depends only on the rotation curve shape, but it is ambiguous for galaxies that have rising rotation curves since the circular velocity may keep increasing beyond the last measured point.
 \item The average circular velocity along the flat part of the rotation curve $V_{\rm f}$. Similarly to $V_{\rm max}$, this velocity definition depends only on the rotation curve shape. We use the automatic algorithm described in \citet{Lelli2016a}, where the flat part is defined by the outer velocity points with relative cumulative differences smaller than 5$\%$.
\end{enumerate}

\begin{figure*}
\centering
\includegraphics[width=0.95\textwidth]{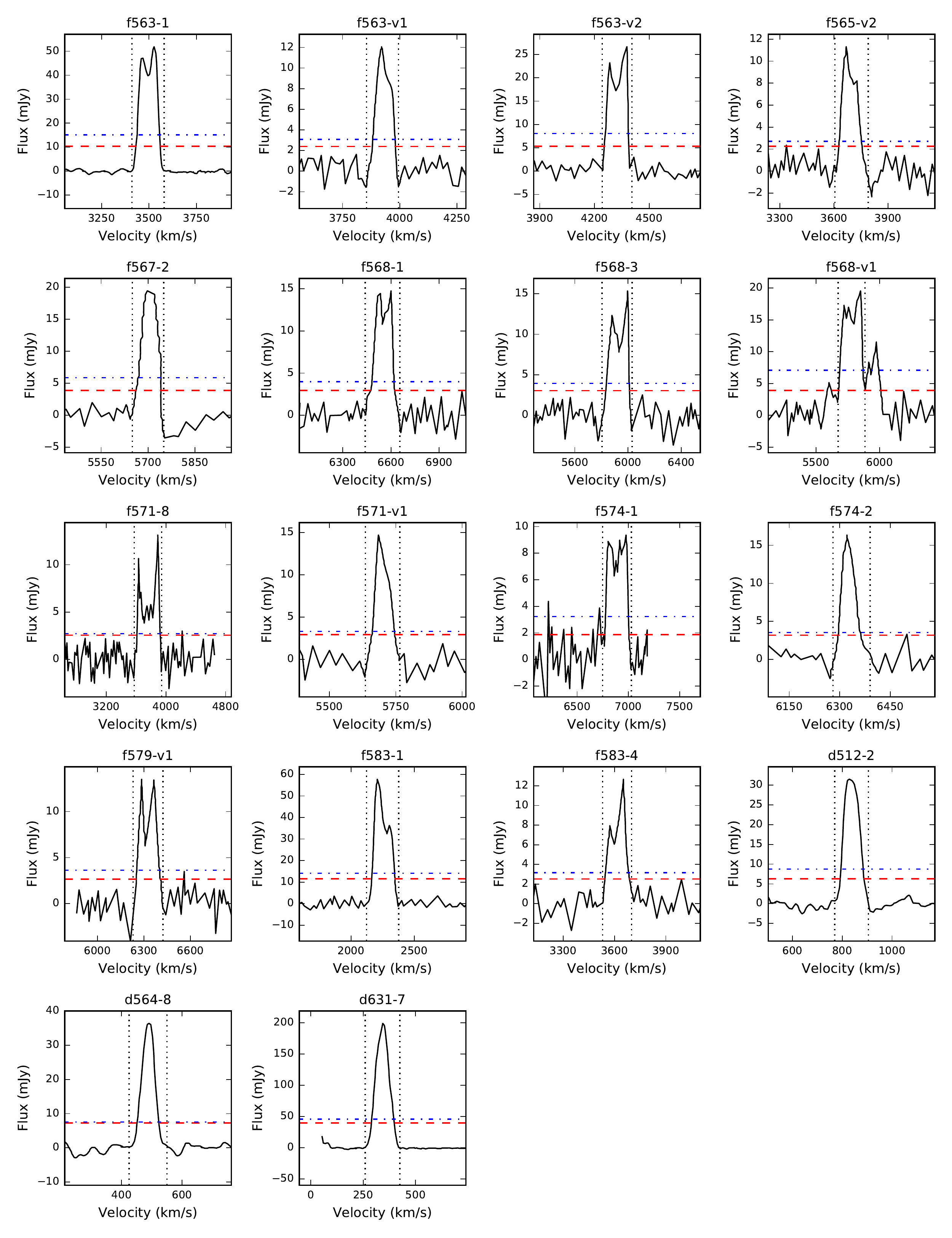}
\caption{Single-dish profiles for 18 LSB galaxies from \citet{Schombert1992} and \citet{Eder2000}. The vertical dotted lines indicate the velocity window used to calculate the mean flux density. The red, dashed and the blue, dash-dotted lines show the flux levels used to calculate \Wp\ and \Wm, respectively.}
\label{fig:HIprof}
\end{figure*}
The velocities (iii) to (vi) are measured using SPARC rotation curves and [3.6] luminosity profiles, as we describe in Sect.\,\ref{sec:rotcur}. The \hi\ line-widths are culled from the literature, as we describe in Sect.\,\ref{sec:singledish}. All these velocity measurements are corrected for disc inclination $i$ adopting the fiducial SPARC values, which mostly come from tilted-ring fits to the velocity fields \citep[see][]{Lelli2016b}. Inclination uncertainties $\delta_{\rm i}$ are propagated on the velocity errors by adding in quadrature the factor $V \cdot \delta_{i}/\tan(i)$. This inclination term typically dominates the error budget, so the final errors for the different velocity definitions are similar and the intrinsic scatters of the different BTFRs can be safely compared.

\subsection{Rotation Curve Data} \label{sec:rotcur}

Rotation curves are typically derived by dividing the galaxy into a series of independent rings, where the ring width is given by the spatial resolution of the observations. To measure \Vd\ and \Ve, we linearly interpolate the sampled points of the rotation curves. In five cases, there is no velocity point within $2.2 R_{\rm d}$. This happens when the \hi\ observations have limited spatial resolution and/or the gas disc has a central hole. A linear interpolation from $R=0$ to the first measured point is not trustworthy, so we provide \Vd\ for 148 out of 153 galaxies. In nine cases, the rotation curve is less extended than $2 R_{\rm e}$. This happens when the \hi\ observations have limited sensitivity and/or the gas disc is truncated. We can measure \Ve\ for 142 out of 153 galaxies.

The errors on \Vd\ and \Ve\ are assigned considering the largest errorbar among the two measured points that bracket \Vd\ and \Ve. In general, these two errorbars are similar, so this choice does not strongly affect our results. To this random error on the circular velocity, we add the inclination term discussed at the end of Sect.\,\ref{sec:VelDef}, which is common to all velocity definitions and dominates the error budget.

We measure \Vf\ for 123 objects. The remaining 30 galaxies do not satisfy the 5$\%$ flatness criteria described in \citet{Lelli2016a} because their rotation curves are either rising or declining in the outer parts. This is likely a matter of sensitivity: deeper \hi\ observations may probe the rotation curve out to larger radii where a flat part is reached (see, e.g., \citealt{Swaters2009}). The 5$\%$ flatness criteria, however, still allows for some residual outer slope. If we consider the innermost and outermost points definying $V_{\rm f}$, the logarithmic slope $S_{1,2} = \log(V_{1}/V_{2})/\log(R_1/R_2)$ has a average value of 0.05 and rms scatter of 0.12. Moreover, the outer slope displays a trend with galaxy luminosity and surface brightness: low-luminosity ($M_\star<10^{10} M_{\odot}$) and low-surface-brightness ($\Sigma_{\rm eff} < 100$ $L_{\odot}$\,pc$^{-2}$) galaxies tend to have rising rotation curves ($S_{1,2}>0$), while high-mass and high-surface-brightness spirals tend to have decling rotation curves ($S_{1,2}<0$).

The errors on \Vf\ are given by Eq.\,3 of \citet{Lelli2016a}, which considers the random errors on the velocity points and the degree of flatness of the rotation curve, other than the disc inclination. However, we revise the values from \citet{Lelli2016a} because we found a minor ``indexing bug'' in our software, which was under-estimating these errors by $\sim$0.6 \kms\ on average (corresponding to 0.1$\%$ difference). This does not affect any of our previous conclusions.

\subsection{Single-dish Data} \label{sec:singledish}

Our primary source of \hi\ line-widths is the extragalactic distance database \citep[EDD,][]{Tully2009}. The EDD contains two major \hi\ catalogs \citep{Courtois2009}: (i) the pre-digital \hi\ catalog provides \Wp\ values that were visually derived by B.\,Tully and collaborators using relatively old single-dish scans, and (ii) the all-digital \hi\ catalog provides \Wm\ values that were derived with a dedicated algorithm using more recent single-dish data. These catalogs provide us with homogeneous measurements of \Wp\ and \Wm\ for 133 and 114 SPARC galaxies, respectively. When multiple estimates of \Wm\ are available in the EDD from different telescopes, we take the arithmetic average. 

The EDD also provides the $W_{\rm mx}$ parameter, in which \Wm\ is corrected for instrumental resolution, relativistic broadening, and turbulent motions. The last correction aims to adjust the observed line-width to twice the maximum rotation velocity and is obtained using \hi\ rotation curves \citep{Courtois2009}. This correction is not adequate for our goals since we aim to compare the BTFR from \hi\ line-widths with that from spatially resolved data, providing new statistical corrections between these different measurements. It may be convenient for us to apply only the corrections for relativistic broadening and instrumental resolution: these are only available for \Wm\ and indicated as \Wmc. For nearby galaxies like those in SPARC, corrections for relativistic broadening are very small and often negligible. Moreover, the errors on line-widths from the EDD are (by construction) larger than the telescope velocity resolution \citep{Courtois2009}, thus instrumental effects are taken into account in the error budget. For the sake of homogeneity, we perform our main analysis on \hi\ line-widths without any correction other than galaxy inclination.

The EDD misses low-surface-brightness (LSB) galaxies from the catalogs of \citet{Schombert1992, Schombert1997}, which are instead well-represented in SPARC. We digitized existing \hi\ profiles from Arecibo observations for 18 LSB galaxies in SPARC (14 from \citealt{Schombert1992} and 4 from \citealt{Eder2000}). We calculated \Wp\ and \Wm\ following similar procedures as \citet{Courtois2009} and assigned a fixed error of 16 km\,s$^{-1}$, which is the velocity resolution of these Arecibo data (after Hanning smoothing). The \hi\ profiles and the recovered \Wp\ and \Wm\ are shown in Figure\,\ref{fig:HIprof}.

For the remaining SPARC galaxies, we search the literature for similar measurements of \Wp\ and \Wm. This introduces some heterogeneity in our \hi\ line-widths, but we prefer to have the largest possible number of datapoints given the relatively small size of SPARC. Different determinations of \Wp\ and \Wm\ from different authors are expected to result in minor differences (within the errors). Specifically, we add the following measurements to the EDD+LSB sample:
\begin{enumerate}
 \item Fourteen measurements of \Wp\ from \citet{Springob2005}. These are comparable to the \Wp\ values from the EDD. Considering the whole EDD pre-digital catalog, \citet{Courtois2009} find that the mean difference between the two measurements is 7$\pm$22 \kms (their Table\,1), which does not introduce any significant systematics.
\item Eight measurements of \Wm\ from \citet{Springob2005}. \citet{Courtois2009} use the notation $W_{\rm m50}$ to distinguish between their determination of the \hi\ line-width at 50$\%$ of the mean flux and the analogous one from \citet{Springob2005}. We do not make this distinction here. \citet{Courtois2009} find that the mean difference between the two determinations is 6.6$\pm$9.0 \kms, indicating that there are no significant systematic differences.
\item Three measurements of \Wp\ from the catalog of \citet{Huchtmeier1989}. This is a compilation of pre-digital line-widths, which are similar in spirit to the \Wp\ values in the EDD. Since errors on \Wp\ are not provided, we consider the velocity resolution of these observations as the minimum error.
\item Two measurements of \Wm\ from the catalog of \citet{Huchtmeier1989}. These are derived in a similar way as \citet{Courtois2009} and \citet{Springob2005}. Again, we consider the velocity resolution of the observations as the minimum error.
\end{enumerate}
In summary, measurements of \Wp\ and \Wm\ are available for, respectively, 168 and 142 galaxies out of 175. Excluding galaxies with low inclination ($i<30$) and low-quality rotation curves ($Q=3$), we have \Wp\ and \Wm\ values for 148 and 125 galaxies out of 153.

\section{Baryonic Tully-Fisher Relations}\label{sec:BTFR}

\begin{figure*}
\centering
\includegraphics[width=0.9\textwidth]{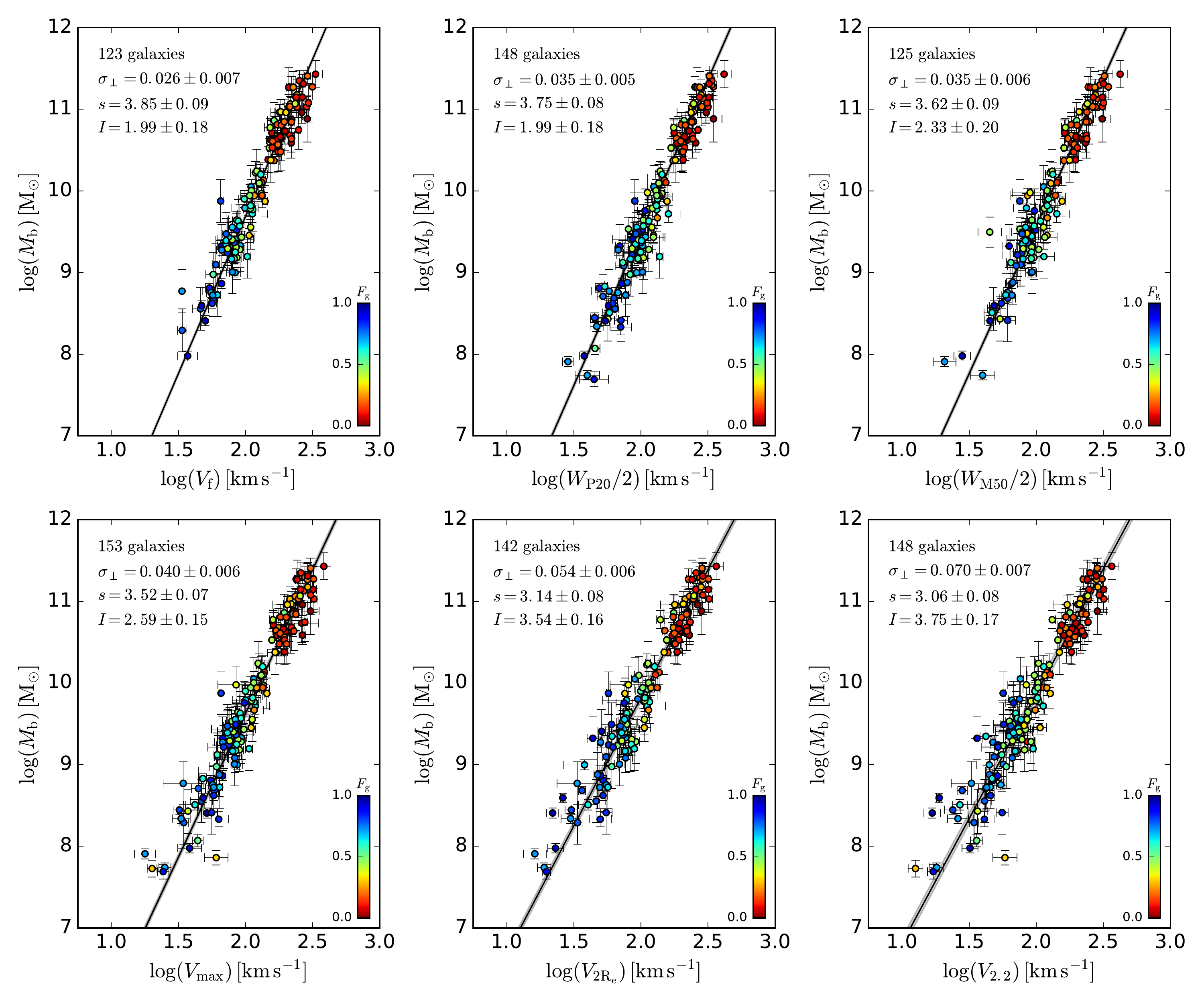}
\caption{The BTFR for different velocity definitions (ordered by increasing intrinsic scatter): \Vf\ (top left), \Wp/2 (top middle), \Wm/2 (top right), \Vm\ (bottom left), \Vd\ (bottom middle), and \Ve\ (bottom right). Points are color-coded by the gas fraction $F_{\rm g} = M_{\rm g}/M_{\rm b}$. The black line shows a linear fit to the data, while the gray shade illustrates the perpendicular intrinsic scatter. The number of galaxies and the best-fit parameters are indicated in the top-left corner. The tightest and steepest BTFR is given by \Vf.}
\label{fig:BTFR}
\end{figure*}

Figure\,\ref{fig:BTFR} shows the BTFR for different velocity definitions. The baryonic mass $M_{\rm b}$ is estimated as
\begin{equation}
 M_{\rm b} = 1.33M_{\hi} + \Upsilon_{\star}L_{[3.6]},
\end{equation}
where $M_{\hi}$ is the \hi\ mass, $L_{[3.6]}$ is the luminosity at [3.6], and $\Upsilon_{\star}$ is the stellar mass-to-light ratio. The factor 1.33 takes the contribution of Helium into account. We neglect the molecular gas, but this generally provides a minor contribution to the total baryonic mass of nearby galaxies \citep[e.g.,][]{Cortese2014, McGaugh2015}. We assume $\Upsilon_{\star} = 0.5$ M$_{\odot}$/L$_{\odot}$ at [3.6] for all galaxies as suggested by both stellar population synthesis models \citep[e.g.,][]{McGaugh2014, Norris2016, Schombert2019} and color-magnitude diagrams of resolved stellar populations \citep[e.g.,][]{Eskew2012, Zhang2017}. 

The errors on $M_{\rm b}$ are estimated using Eq.\,5 of \citet{Lelli2016a}, which considers uncertainties on total fluxes and galaxy distances, as well as galaxy-to-galaxy variations in $\Upsilon_{\star}$ with a standard deviation of 0.11 dex. To have a direct comparison with \citet{Lelli2016a}, we do not treat bulge and disc separately, assigning different values of $\Upsilon_{\star}$. This would give slightly steeper BTFR slopes because bulges have higher $\Upsilon_{\star}$ than discs and are predominantly found at high $M_{\rm b}$. The effects of different values of $\Upsilon_{\star}$ on the BTFR are discussed in \citet{Lelli2016a} and \citet{Li2018}. Here we focus on the effects of different velocity definitions for a fixed definition of $M_{\rm b}$.

\subsection{MCMC fitting}\label{sec:MCMC}
\begin{table}
\caption{BTFRs for different velocity definitions ($x$). We fit the linear relation $\log(M_{\rm b}/M_{\odot}) = s\,\log(x/{\rm km\,s^{-1}}) + I$, considering errors on both variables and orthogonal intrinsic scatter ($\sigma_\perp$). The last column gives the number of galaxies in each sub-sample.}
\begin{center}
\setlength{\tabcolsep}{5.0pt}
\begin{tabular}{l c c c c c}
\hline
$x$    & $s$ & $I$ & $\sigma_{\perp}$ & $\sigma_{\rm o}$ & $N$\\
\hline
\Vf\   & 3.85$\pm$0.09 & 1.99$\pm$0.18 & 0.026$\pm$0.007 & 0.24 &123 \\
\Wp/2  & 3.75$\pm$0.08 & 1.99$\pm$0.18 & 0.035$\pm$0.005 & 0.24 &148 \\
\Wm/2  & 3.62$\pm$0.09 & 2.33$\pm$0.20 & 0.035$\pm$0.006 & 0.27 &125 \\
\Vm\   & 3.52$\pm$0.07 & 2.59$\pm$0.15 & 0.040$\pm$0.006 & 0.27 &153 \\
\Ve\   & 3.14$\pm$0.08 & 3.54$\pm$0.16 & 0.054$\pm$0.006 & 0.26 &142 \\
\Vd\   & 3.06$\pm$0.08 & 3.75$\pm$0.17 & 0.070$\pm$0.007 & 0.30 &148 \\
\hline
\end{tabular}
\end{center}
\label{tab:BTFR}
\end{table}
We perform linear fits of the form:
\begin{equation}
 \log(y) = s\,\log(x) + I,
\end{equation}
where $y=M_{\rm bar}$ in $M_{\odot}$, $x$ is one of the velocity definitions in \kms, $s$ and $I$ are free parameters. It is not trivial to fit a line to data points with errors on both variables as well as intrinsic scatter. Indeed, it has been argued that the properties of the BTFR may depend on the fitting technique \citep{Bradford2016}. In Appendix \ref{sec:fitting}, we describe three different fitting methods: (1) the generalized least-squares method \citep[as in][]{Lelli2016a}, (2) the maximum likelihood method assuming a Gaussian model with constant intrinsic scatter along the vertical direction ($\sigma_y$), and (3) the maximum likelihood method assuming a Gaussian model with constant intrinsic scatter along the orthogonal direction to the best-fit line ($\sigma_\perp$). These fitting methods give consistent results within the errors, albeit there are subtle systematic differences that we describe in Appendix \ref{sec:fitting}. The major differences in the BTFR properties are driven by the choice of the velocity definition, not by the choice of the fitting method.

For our fiducial analysis, we use the maximum likelihood method that considers the orthogonal intrinsic scatter. This likelihood function is more concentrated around the best-fit line than the others, giving the data a higher maximum likelihood, so it is preferred in a model comparison sense. We run Markov Chain Monte Carlo (MCMC) simulations to map the posterior probability distributions of $s$, $I$, and $\sigma_\perp$. We use the standard affine-invariant ensemble sampler in $emcee$ \citep{Foreman-Mackey2013} and initialize the chains with 50 random walkers. The start position of the walkers is randomly set within realistic ranges: [3.0, 5.0] for $s$, [1.0, 5.0] for $I$, and [0.01, 0.25] for $\sigma_\perp$. We run 1000 burnt-in iterations and re-run the simulation with 1000 steps. We visually checked the convergence of the chains.

Table\,\ref{tab:BTFR} provides the best-fit parameters and associated errors for the various velocity definitions. We also report the number $N$ of galaxies in each sub-sample and the vertical observed scatter $\sigma_{\rm o}$ given by:
\begin{equation}
 \sigma_{\rm o} = \sqrt{ \dfrac{1}{N}\sum^{N}\bigg[\log(M_{\rm b}) - s\,\log(x) - I\bigg]^2 }.
\end{equation}
The errors on the best-fit parameters are estimated as the standard deviation around the maximum-likelihood value in the posterior distributions. These errors are indicative because the posterior distributions are not exactly Gaussian. Figure\,\ref{fig:corner} shows the posterior parameter distributions for the $M_{\rm b}-\Vf$ relation as an example. In this case, we have $\sigma_{\perp}=0.026\pm0.007$ dex but the posterior distribution has a non-Gaussian tail towards low values, so $\sigma_{\perp}$ is actually consistent with zero within the 3$\sigma$ confidence region.

\subsection{Observed and intrinsic scatter}\label{sec:scatter}

The tightest BTFR is given by \Vf\ in terms of both observed and intrinsic scatters. The broadest relation is given by \Vd. This is somewhat counter-intuitive as \Vd\ is measured in the inner galaxy regions where baryons are dynamically important, whereas \Vf\ is measured in the outermost regions where dark matter usually dominates. Therefore, one may expect that $M_{\rm b}$ is more closely related to \Vd\ rather than \Vf. This does not occur in Nature. The most fundamental BTFR links $M_{\rm b}$ and \Vf. This is in line with previous works on smaller galaxy samples \citep{Verheijen2001b, McGaugh2005, McGaugh2005b, Noordermeer2007b, Ponomareva2018}. 

Interestingly, \hi\ line-widths give significantly tighter BTFRs than \Vm, \Ve, or \Vd. Hence, it is better to have a spatially integrated measurement in the outer regions than a spatially resolved one in the inner parts. In other words, it is preferable to sacrifice the precision of the velocity measurement in favour of its radial extent. \Wp\ and \Wm\ display nearly the same values of $\sigma_{\perp}$, but \Wp\ returns steeper slopes than \Wm. This may indicate that \Wp\ probes slightly larger radii than \Wm\ (see Sect.\,\ref{sec:slope}).

We stress that all these velocities have been corrected for disc inclination using best-fit values from \hi\ velocity fields. Single-dish observations can provide \Wp\  and \Wm\ for large galaxy samples, but carry no information on $i$. In such single-dish \hi\ surveys, disc inclinations are typically estimated using optical images and can be significantly more uncertain than the ones adopted here, potentially adding significant scatter in the BTFR from \hi\ line-widths.

The \Vm\ values give intermediate results in terms of scatter. This is likely due to the ill-defined nature of \Vm. For bulge-dominated galaxies with declining rotation curves in the inner parts, \Vm\ occurs at very small radii and we have $\Vm  > \Vf$. For disc-dominated galaxies with perfectly flat rotation curves, \Vm\ occurs at intermediate radii and we have $\Vm \simeq \Vf$. For some dwarf galaxies with rising rotation curves, \Vm\ occurs at the last measured point and we have $\Vm < \Vf$ since the flat part has not been reached. It is likely that these rising rotation curves are merely a matter of sensitivity: deeper \hi\ observations could probe the circular velocities out to larger radii, where the rotation curve presumably flattens out. The inclusion of these rising rotation curves on the BTFR (as done with \Vm) artificially increases its scatter and introduces severe systematics on slope and normalization, as we discuss in the next Section.

One may wonder whether the BTFR from $V_{\rm f}$ is tighter than that from other velocity measurements because of the smaller galaxy sample. Indeed, our criteria on rotation curve flatness selects 123 galaxies out of 153. We repeated the same analysis considering a sub-sample of 99 galaxies for which all velocity measurements are simultaneously available. The best-fit results display the same general trends. Moreover, the results from the sub-sample are consistent with those from the full sample within 1.5$\sigma$, indicating that slope, intercept, and intrinsic scatter are robust against minor modifications of the galaxy sample.

\subsection{Slope and intercept}\label{sec:slope}

\begin{figure}
\centering
\includegraphics[width=0.45\textwidth]{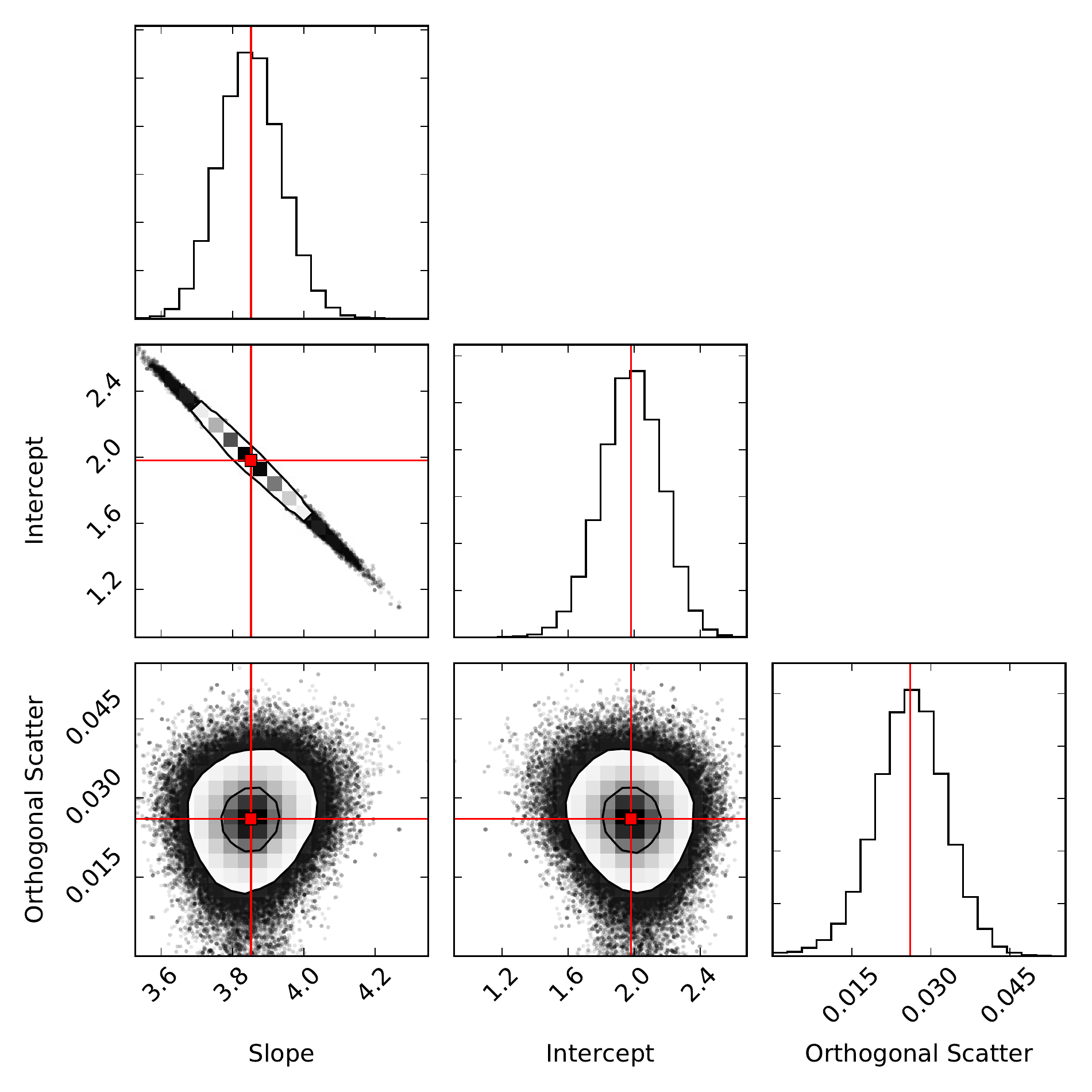}
\caption{Posterior distributions of the fitting parameters in the case of \Vf, which provides the tightest BTFR. Red squares and solid lines indicate the maximum-likelihood values. Black contours correspond to 1$\sigma$ and 2$\sigma$ confidence regions.}
\label{fig:corner}
\end{figure}
The BTFR from \Vf\ has the smallest intrinsic scatter, so we consider its slope and normalization as the fiducial ones. Using the orthogonal maximum-likelihood method, we find $s = 3.85\pm0.09$ (see Fig.\,\ref{fig:corner} for the full posterior distribution). The slope is even steeper if one weights galaxies by their gas fraction, minimizing the effect of $\Upsilon_{\star}$ \citep{Lelli2016a}, or if one treats bulge and disc separately, assigning higher $\Upsilon_{\star}$ to bulges \citep{Li2018}. This is important because different theories predict different BTFR slopes: (i) the simplest $\Lambda$CDM models with constant baryonic fractions ($M_{\rm b}/M_{\rm halo}$) predict a slope of $\sim$3 \citep{Mo1998, McGaugh2012}, (ii) recent $\Lambda$CDM models with baryonic fractions from abundance matching give steeper slopes from $\sim$3.5 to $\sim$3.7 \citep{DiCintio2016, Desmond2017, Sales2017}, and (iii) modified Newtonian dynamics \citep[MOND, ][]{Milgrom1983} predicts a slope of 4 (formally this holds for $V \rightarrow V_{\rm f}$ as $R\rightarrow \infty$ for completely isolated galaxies).

\begin{figure*}
\centering
\includegraphics[width=0.95\textwidth]{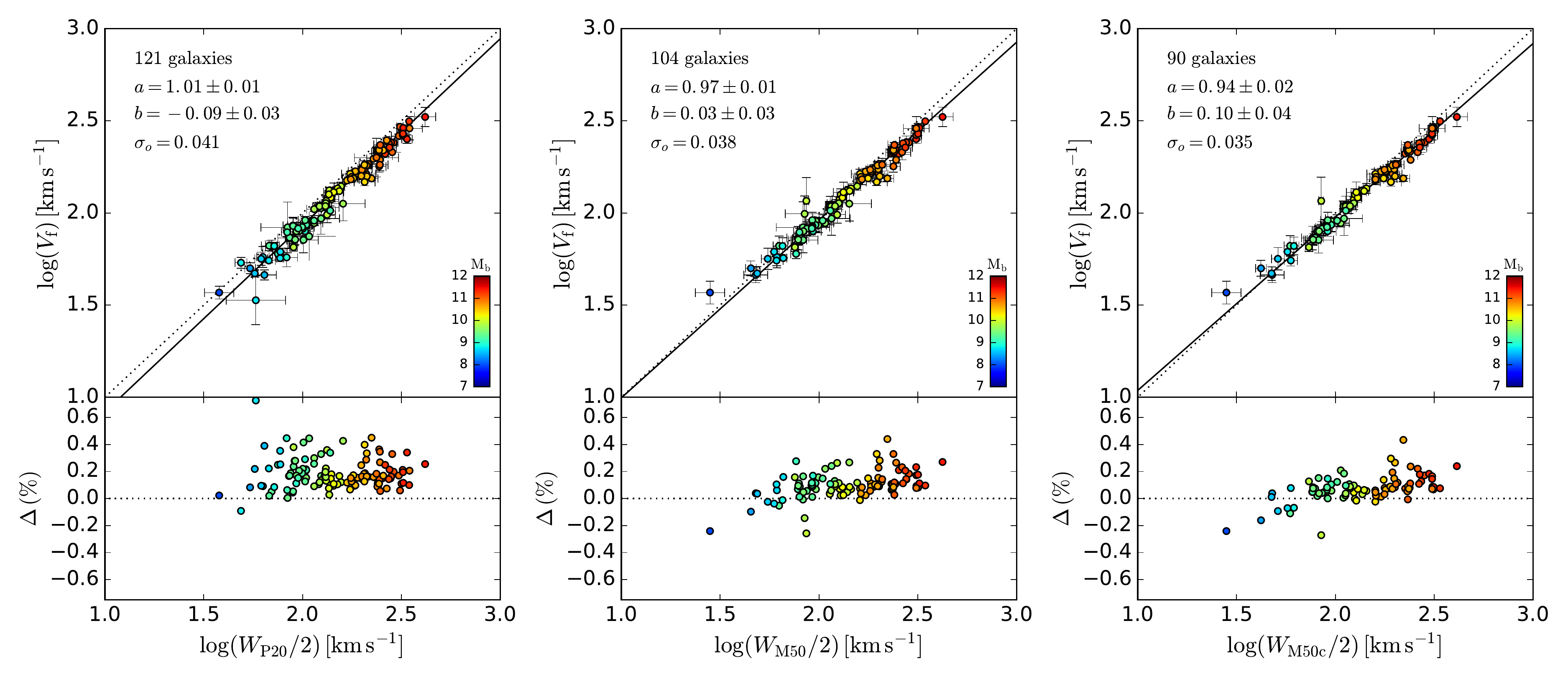}
\includegraphics[width=0.95\textwidth]{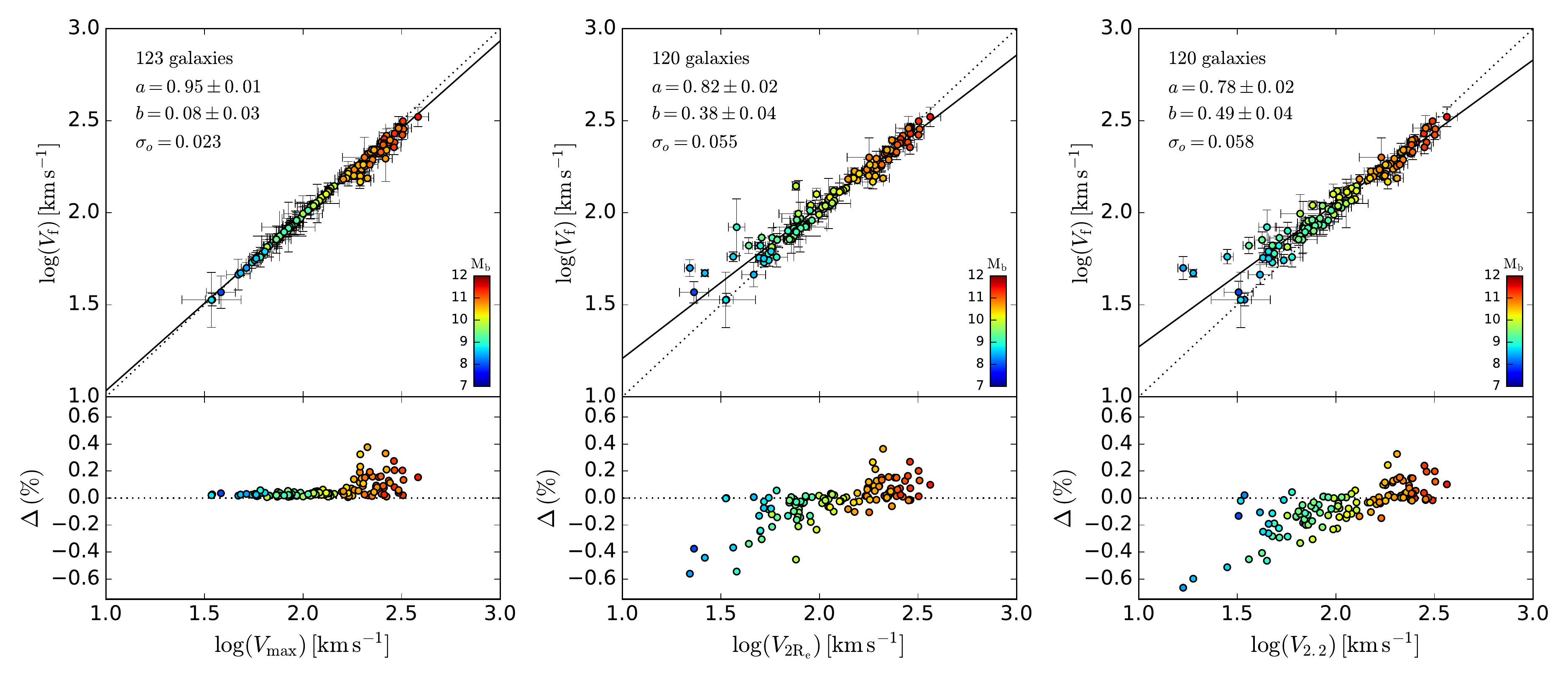}
\caption{Relations between $V_{\rm f}$ and other velocity definitions: $W_{\rm P20}/2$ (top left), $W_{\rm M50}/2$ (top middle), $W_{\rm M50c}/2$ (top right), $V_{\rm max}$ (bottom left), $V_{\rm 2.2}$ (bottom middle), and $V_{\rm eff}$ (bottom right). The subplots show relative differences $\Delta = (x - \Vf)/\Vf$, where $x$ is the velocity definition along the abscissa. Galaxies are color-coded by baryonic mass. The dotted line corresponds to equality. The solid line shows a linear fit to the data. The number of galaxies and the best-fit parameters are indicated in the top-left corner.}
\label{fig:Vcor}
\end{figure*}

All the other velocity definitions give systematically shallower slopes than \Vf. This corresponds to higher BTFR normalizations, given the strong degeneracy between $s$ and $I$ (Fig.\,\ref{fig:corner}). These systematic effects are well understood: they are due to the variation of rotation curves shapes with baryonic mass \citep[e.g.,][]{Verheijen2001b}. In general, high-mass galaxies have rotation curves that rise very fast at small radii, decline significantly at intermediate radii, and become flat in the outer parts. On the other hand, low-mass galaxies have slowly rising rotation curves that reach a flat part only in the outermost regions. Velocity definitions probing inner radii give rotation velocities systematically higher than \Vf\ for high-mass galaxies, moving these galaxies to the right of the fiducial BTFR, but systematically smaller for low-mass galaxies, moving these galaxies to the left of the fiducial BTFR. The net result is a relation with a systematically shallower slope. The systematics between different velocity definitions are further discussed in the next section.

We recall that the global \hi\ profile of a galaxy is a spatial convolution between the galaxy rotation curve and the \hi\ surface density profile. Thus, although \hi\ line-widths probe the rotation velocity at larger radii than \Vd, they are still prone to systematic effects in determining the BTFR slope and normalization \citep{Papastergis2016}. For example, several low-mass galaxies reach \Vf\ in the outermost regions where the \hi\ surface densities are very low ($\sim$10$^{19}$ atoms\,cm$^{-2}$) and do not contribute significantly to the spatially integrated \hi\ flux. Hence, although they do reach a flat part, their global \hi\ profiles will not show the classic ``double-horned'' shape seen in spiral galaxies and their \hi\ line-width will under-estimate \Vf\ to some extent. Similarly, the global \hi\ profiles of high-mass galaxies are generally skewed towards high-density gas in the inner regions, where the rotation curve is declining towards \Vf. Hence, even if their global \hi\ profiles may show a ``double-horned'' shape, their \hi\ line-width may over-estimate \Vf. Thus, \hi\ line-widths give systematically shallower slopes than \Vf.

In terms of \hi\ line-widths, our results are directly comparable to \citet{Zaritsky2014} as they exploit \Wm\ measurements from the EDD and stellar masses from $Spitzer$ photometry, adopting a similar stellar-mass calibration as in our work \citep{Eskew2012}. \citet{Zaritsky2014} select 903 galaxies from a magnitude-limited, volume-limited parent sample and report a slope of $3.5\pm0.2$, which is fully consistent with our slope of $3.62\pm0.09$ for the same velocity definition. This confirms that the SPARC sample is representative for the general population of disc galaxies.

For a fixed velocity definition, the remaining systematics on the BTFR slope are the choice of the mean $\Upsilon_\star$ \citep[see][]{Lelli2016a} and the fitting method (see Appendix\,\ref{sec:fitting}). The stellar population models of \citet{Schombert2019} give mean $\Upsilon_\star$ at [3.6] between 0.4 and 0.6 for star-forming discs, depending on the adopted galaxy formation scenario and treatment of asymptotic giant branch stars (see their Figure 11). Bulges likely have higher $\Upsilon_\star$ values, up to 0.7$-$0.8 at [3.6]. Considering all these systematic uncertainties, the BTFR from \Vf\ has a slope in the range 3.5 to 4.0.

\section{Statistical corrections between different velocity definitions}\label{sec:sys}

As we discussed in Sect.\,\ref{sec:BTFR}, $V_{\rm f}$ gives a tighter and steeper BTFR than other velocity definitions. These systematic differences can be worrisome in several circumstances. For example, rotation curves of high-$z$ galaxies can usually be traced out to smaller radii than those of local galaxies because (1) ionised and molecular gas are the only available tracers to date, and (2) long telescope exposure times are needed due to cosmological surface brightness dimming. Thus, it is not always possible to reach $V_{\rm f}$ at high $z$ and have a proper comparison with the BTFR at $z\simeq0$. Another example is the prospect of studying the environmental dependence of the BTFR using large galaxy samples with \hi\ line-widths: the slope of the $M_{\rm b}-W_{\hi}$ relation could depend on the ratio between low and high mass galaxies in specific environments, but it would not necessarily imply a real physical variation in the more fundamental $M_{\rm b}-V_{\rm f}$ relation.

Figure\,\ref{fig:Vcor} displays the relations and relative differences between \Vf\ and other velocity definitions. Systematic effects are evident. The most problematic cases are \Ve\ and \Vd: high-mass (low-mass) galaxies have declining (rising) rotation curves in their inner parts, so \Ve\ and \Vd\ are progressively higher (lower) than $V_{\rm f}$. For \Vm\ this effect occurs only at the high-mass end because (by construction) $\Vf \simeq \Vm$ for galaxies without inner declines in their rotation curves.

\Wp\ and \Wm\ show systematic shifts with respect to \Vf. Indeed, the BTFR from $W_{\rm P20}$ has similar slope as the BTFR from $V_{\rm f}$ but a lower normalization. The vast majority of $W_{\rm P20}$ data are taken from the pre-digital \hi\ archive of the EDD. They come from relatively old single-dish observations (dating back to the 1970s) and were not corrected for the instrumental resolution, which was coarse in the early days of radio astronomy. It is likely that the systematic shift with respect to $V_{\rm f}$ is due to the low spectral resolution of pre-digital observations, broadening the global \hi\ line profiles. Unfortunately, we do not have the necessary information to correct $W_{\rm P20}$ for instrumental resolution. However, we can have a rough idea of the magnitude of this correction by comparing the values of $W_{\rm M50}$ (uncorrected) and $W_{\rm M50c}$ (corrected) from the all-digital \hi\ archive of the EDD. The $V_{\rm f}- W_{\rm M50}$ and $V_{\rm f} - W_{\rm M50c}$ relations have comparable slopes but display a systematic shift in normalization of 0.1 dex. This is similar to what we need to make $W_{\rm P20}$ consistent with $V_{\rm f}$. 

We perform linear fits of the form:
\begin{equation}
 \log(y) = a\,\log(x) + b,
\end{equation}
where $y=V_{\rm f}$ and $x$ is one of the other velocity definitions (both in \kms). These relations can be used to statistically infer \Vf\ from less extended rotation curve data (\Vd\ or \Ve) or spatially unresolved observations (\Wp\ or \Wm). Thus, it may be possible to reconstruct the $M_{\rm b}-\Vf$ relation in the absence of extended \hi\ rotation curves. Since we are interested in the random variation of \Vf\ for a given $x$, we fit for the vertical intrinsic scatter $\sigma_{V_{\rm f}}$ using the generalized least-squares method (described in Appendix \ref{sec:fitting}). The results are given in Table\,\ref{tab:Vcor}. The \textit{observed} scatter in the $V_{\rm f} - x$ relations is much smaller than the \textit{intrinsic} scatter in the respective $M_{\rm b} - x$ relations, thus these statistical corrections will not add too much noise in the final $M_{\rm b} - V_{\rm f}(x)$ relation. The observed scatter provides a conservative estimate of the error in converting a given velocity definition into $V_{\rm f}$.

\begin{figure*}
\centering
\includegraphics[width=0.95\textwidth]{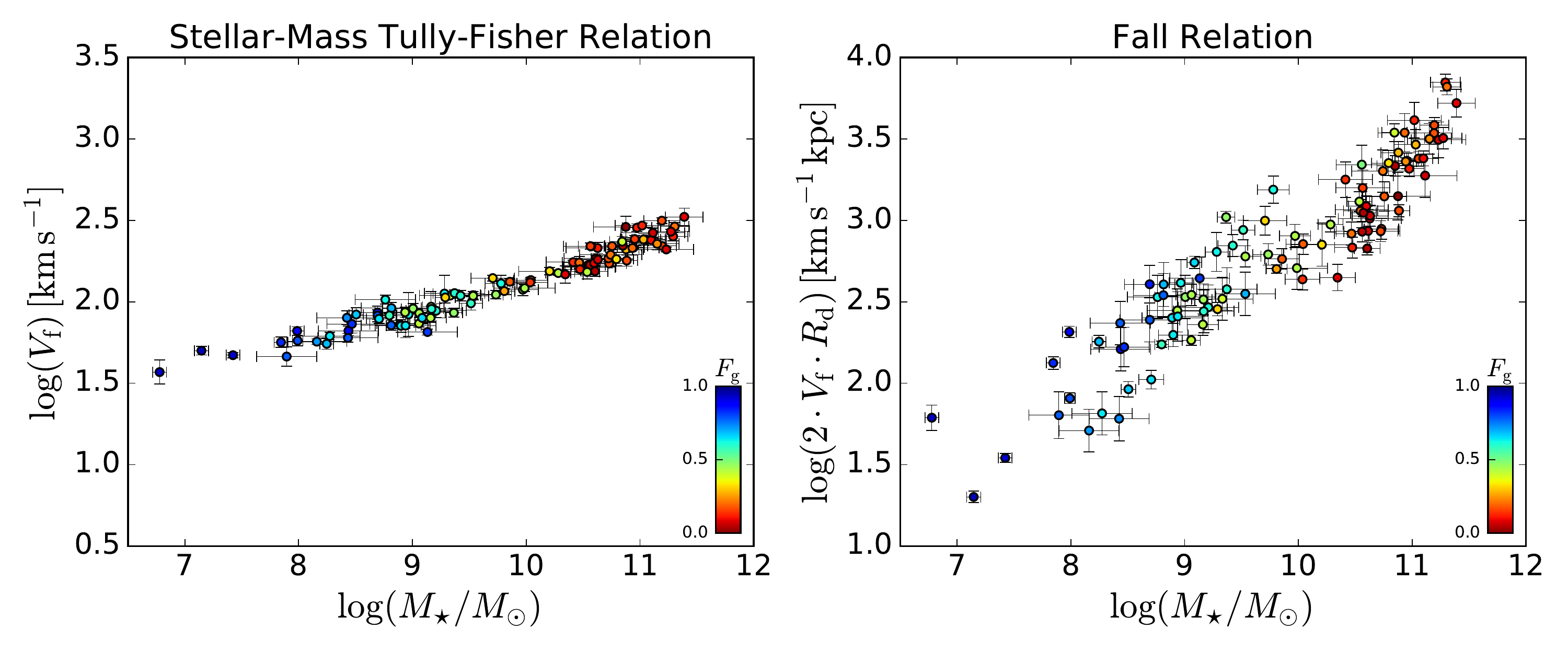}
\caption{Comparing the stellar-mass TF relation (left) with the Fall relation (right). In both panels, the vertical axis spans 3 orders of magnitude, so one can visually compare the tightness of the two relations. The stellar-mass TF relation is $\sim$5 times tighter than the Fall relation. The additional scatter in the Fall relation is entirely driven by the scatter in the mass-size relation of galaxies.}
\label{fig:TFvsFall}
\end{figure*}

\begin{table}
\caption{Relations between \Vf\ and other velocity definitions ($x$). We fit the linear relation $\log(V_{\rm f}) = a\,\log(x) + b$, considering intrinsic vertical scatter ($\sigma_{V_{\rm f}}$). The last two columns give the observed vertical scatter and the number of datapoints in each sample.}
\begin{center}
\setlength{\tabcolsep}{3pt}
\begin{tabular}{l c c c c c}
\hline
x              & a             & b     & $\sigma_{V_{\rm f}}$ & $\sigma_{\rm o}$ & $N$\\
\hline
$W_{\rm P20}/2$  & 1.01$\pm$0.01 & -0.09$\pm$0.03 & $<$0.003 & 0.041 & 121 \\
$W_{\rm M50}/2$  & 0.96$\pm$0.01 & 0.03$\pm$0.03 & -                 & 0.038 & 104 \\
$W_{\rm M50c}/2$ & 0.94$\pm$0.01 & 0.10$\pm$0.03 & -                & 0.035 & 90 \\
$V_{\rm max}$  & 0.95$\pm$0.01 & 0.08$\pm$0.03 & -                & 0.023 & 123 \\ 
$V_{\rm 2R_e}$  & 0.82$\pm$0.02 & 0.38$\pm$0.04 & 0.034$\pm$0.005  & 0.055 & 120\\  
$V_{\rm 2.2}$  & 0.78$\pm$0.02 & 0.49$\pm$0.03 & 0.037$\pm$0.005  & 0.058 & 120 \\
\hline
\end{tabular}
\end{center}
\label{tab:Vcor}
\end{table}
The SPARC sample spans very broad ranges in stellar mass (from $\sim10^7$ to $\sim3 \times 10^{11}$ M$_{\odot}$), effective stellar surface density (from a few M$_\odot$ pc$^{-2}$ to $\sim$3000 M$_\odot$ pc$^{-2}$), and optical morphologies (from S0/Sa to dwarf irregulars), so these corrections have general validity for galaxies in the field and group environments. We warn, however, that they may be unreliable for face-on galaxies ($i<30^{\circ}$), which should be anyhow excluded in BTFR studies. If we aim to apply these corrections to high-$z$ galaxies, we must assume that the shapes of galaxy rotation curves do not evolve significantly with $z$.

\section{Lack of residual correlations and specific angular momentum}\label{sec:residuals}

Broadly speaking, a galaxy can be described by three basic parameters: its total mass $M$, its characteristic velocity $V$, and its characteristic radius $R$. These three parameters appear in physically relevant quantities, such as the total angular momentum with a characteristic value of $M\cdot V\cdot R$. It is interesting, therefore, to investigate whether the residuals around different $M-V$ relations correlate or not with different definitions of $R$.

We study the BTFR residuals versus galaxy effective radius $R_{\rm eff}$ and disc scale length $R_{\rm d}$, using the Pearson's, Spearman's, and Kendall's tests. In all cases, there are no statistically significant correlations: the correlation coefficients are always within $\pm$0.2. A correlation between BTFR residuals and $R_{\rm eff}$ was advocated by \citet{Zaritsky2014}, using $W_{\rm M50c}$ values from the EDD for a large sample of galaxies. For their preferred BTFR slope of 3.5, this putative correlation appears very weak (see the right panel of their Fig.\,4). This residual correlation is not found in our data. In \citet{Desmond2018}, we investigate residual correlations in more detail, considering velocity measurements at several multiples of $R_{\rm eff}$ and performing a detailed comparison with abundance-matching models in a $\Lambda$CDM cosmology.

In the past years, there has been a renewed interest in the relation between specific angular momentum $j \simeq V \cdot R$ and stellar mass $M_{\star}$ of galaxies, hereafter the ``Fall relation'' \citep{Fall1983, Romanowsky2012, Posti2018, Fall2018}. The lack of correlation between BTFR residuals and $R$ suggests that introducing a radial variable in the $M-V$ plane will unavoidably degradate the original relation and increase its scatter, since there is significant intrinsic scatter in the $M-R$ relation of galaxies (see, e.g., Figure 2 in \citealt{Lelli2016b}).

This is explicitly shown in Figure\,\ref{fig:TFvsFall}, where we plot the stellar-mass TF relation and the Fall relation on the same logarithmic scale. Following \citet{Romanowsky2012}, we define $j = 2\cdot V_{\rm f}\cdot R_{\rm d}$, which is appropriate for late-type disc galaxies. It is clear that the Fall relation has much more scatter than the stellar-mass TF relation. A linear fit returns a vertical intrinsic scatter of 0.16 dex ($\sim37 \%$) for the Fall relation and 0.029 ($\sim7\%)$ for the stellar-mass TF relation\footnote{For comparison with previous works on the Fall relation, the axes of the stellar-mass TF relation in Figure\,\ref{fig:TFvsFall} have been swapped with respect to those in Figure\,\ref{fig:BTFR}, so this vertical intrinsic scatter $\sigma_{\rm V_f}$ is not the same quantity as $\sigma_{\rm M_b}$ from, e.g., \citet{Lelli2016a}.}. Hence, the Fall relation is $\sim5$ times broader than the stellar-mass TF relation. This broadening seems entirely driven by the vertical scatter in the mass-size relation of galaxies, which is between 0.1 and 0.2 dex in late-type galaxies \citep{Wu2018}. Similar results are found considering different definitions of $V$ and/or $R$ (like the effective radius) and/or $M$ (like the baryonic mass).

Recently, \citet{Posti2018b} studied the Fall relation using the SPARC sample. They estimated the stellar specific angular momentum $j_\star$ by integrating over the SPARC rotation curves and considering the asymmetric drift correction. This provides a more precise measurement of $j_\star$ than simply $2\cdot V_{\rm f}\cdot R_{\rm d}$. The overall picture, however, does not change. \citet{Posti2018b} report an orthogonal intrinsic scatter of 0.17$\pm$0.01 dex ($\sim$40$\%$) in the Fall relation, which is $\sim$7 times larger than that in the BTFR ($\sim$6$\%$). Regardless of the precise definitions of $M$, $V$, $R$ or $j$, the TF relation must be more fundamental than the Fall relation because it is tighter and displays no room for additional variables.

One may argue that the Fall relation has the advantage of including late-type galaxies and early-type galaxies in the same diagram. In our opinion, this is misleading because early-type galaxies follow the same BTFR as late-type galaxies \citep{denHeijer2015}, providing that one can measure $V_{\rm f}$ from extended \hi\ discs. The BTFR is not a matter of galaxy morphology, gas content, or star-formation activity; the BTFR is simply a matter of being able to measure the circular velocity in the outermost galaxy regions.

Considering simultaneously the BTFR and the mass-size relation of galaxies provides a stronger constraint on galaxy formation models than considering the Fall relation alone. One needs to explain the significant intrinsic scatter in the mass-size plane (possibly driven by intrinsic variations in the angular momentum), while simultaneously reproducing the extremely small intrinsic scatter in the BTFR. This is not trivial to achieve in standard $\Lambda$CDM models \citep{Desmond2015, Desmond2017}. This unexpected phenomenology was predicted by MOND \citep{Milgrom1983}.

\section{Conclusions}\label{sec:Conc}

We studied the BTFR considering different measures of characteristic velocities. \hi\ and H$\alpha$ rotation curves from the SPARC database were used to estimate the mean velocity along the flat part of the rotation curve (\Vf), the maximum velocity (\Vm), the velocity at 2.2 disc scale lengths (\Vd), and the velocity at two effective radii (\Ve). Global \hi\ line-widths (\Wp\ and \Wm) from single-dish observations were either culled from the literature or measured from existing data. Our main results can be summarized as follows:
\begin{itemize}
 \item The tightest BTFR is given by \Vf. Assuming $\Upsilon_\star=0.50\pm0.13$ $M_\odot/L_\odot$, a Gaussian likelihood model gives an orthogonal intrinsic scatter of $0.026 \pm 0.007$ dex ($\sim$6$\%$) and a slope of $3.85\pm0.09$. The BTFR intrinsic scatter is extremely small for astronomical standards and would be even smaller if the galaxy-to-galaxy scatter in $\Upsilon_\star$ is larger than assumed (0.11 dex). The precise value of the slope depends on the fitting method and on the choice of $\Upsilon_\star$ \citep{Lelli2016a}, but is constrained to be between 3.5 and 4.0.
 \item Different velocity definitions give systematically higher scatters and shallower slopes. Interestingly, \hi\ line-widths give tighter BTFR than \Vd, \Ve, or \Vm, indicating that approximate velocity measurements in the outer parts are preferable to precise measurements in the inner regions. 
 \item We provide statistical corrections to infer \Vf\ from spatially unresolved data (like single-dish \hi\ observations) or less extended rotation curves (like H$\alpha$ and CO observations). These can be useful to study the possible environmental dependence of the BTFR with large up-coming \hi\ surveys, as well as to investigate its possible cosmic evolution with ionised and molecular gas surveys at high $z$.
 \item For all velocity definitions, the residuals show no significant correlation with galaxy radius. This implies that the BTFR is superior to the relation between specific angular momentum and galaxy mass (the Fall relation), where an additional radial variable is included. The Fall relation from \citet{Posti2018b}, indeed, has $\sim$7 times more intrinsic scatter than the BTFR. This appears to be fully driven by the scatter in the mass-size relation of galaxies. The BTFR is already the ``fundamental plane'' of galaxy disks: no value is added with a radial variable as a third parameter.
\end{itemize}



\begin{appendix}

\section{Fitting methods}\label{sec:fitting}

\begin{figure*}
\centering
\includegraphics[width=0.9\textwidth]{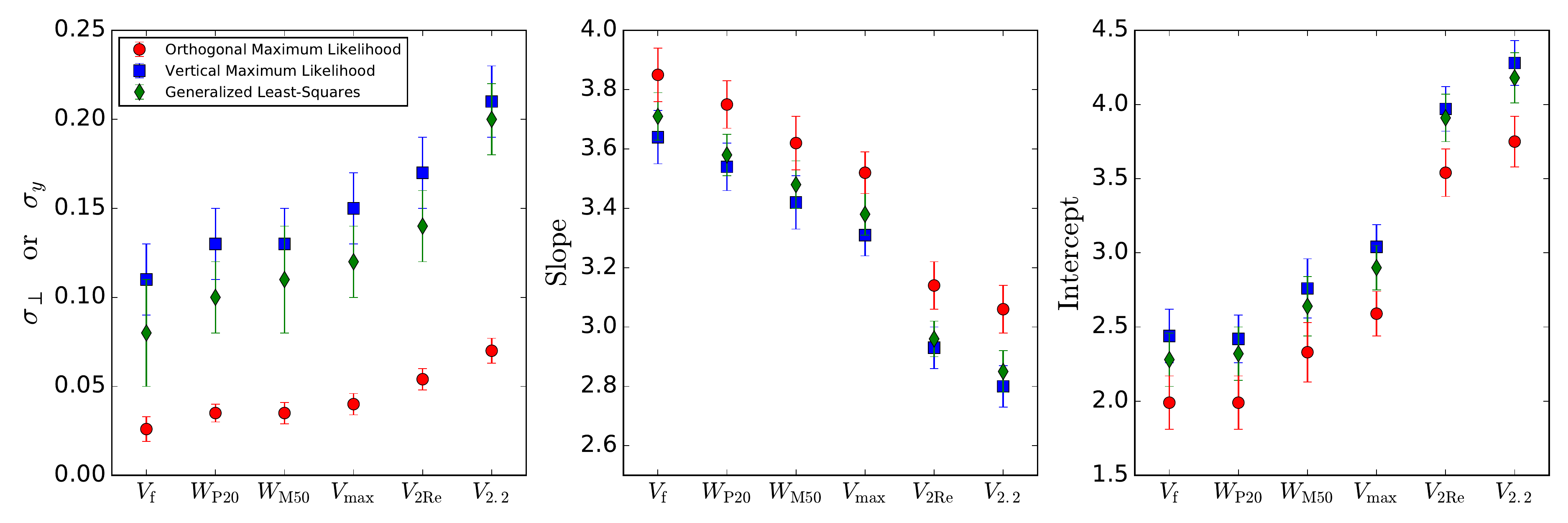}
\caption{The BTFR properties for different fitting methods and velocity definitions. The left panel shows $\sigma_\perp$ for the orthogonal ML method, as well as $\sigma_y$ for the GLS and the vertical ML methods. The middle and right panels show, respectively, the best-fit slope and intercept. Different fitting methods give consistent results within 1.5$\sigma$. The major systematic differences between BTFR properties are driven by the velocity definition.}
\label{fig:Fitting}
\end{figure*}
We describe three common statistical methods to fit a linear model $y = ax +b$ to a set of $N$ points $(x_i, y_i)$ with errors $(e_{xi}, e_{yi})$ and non-negligible intrinsic scatter. These methods are (1) the generalized least-squares (GLS) method; (2) the maximum likelihood method (ML) assuming a Gaussian model with constant intrinsic scatter along the vertical direction ($\sigma_y$), hereafter ``vertical ML''; and (3) the maximum likelihood method assuming a Gaussian model with constant intrinsic scatter along the orthogonal direction to the best-fit line ($\sigma_\perp$), hereafter ``orthogonal ML''.

The GLS method generalizes the standard least-square method by adding the vertical intrinsic scatter $\sigma_y$ in quadrature to the errors \citep[e.g.,][]{Press1992}. We define
\begin{equation}\label{eq:GLS}
 \chi^2 = \sum_{i}^{N} \dfrac{\Delta_{yi}^2}{\delta_{y,\,i}^2},
\end{equation}
where
\begin{equation}\label{eq:ydist}
\Delta_{y,\,i} = y_i - (a x_i + b),
\end{equation}
and
\begin{equation}\label{eq:sigma}
\delta_{y, i} = \sqrt{a^2 e^2_{xi} + e^2_{yi} + \sigma^2_y}.
\end{equation}
The best-fit values are inferred by requiring that $\chi^2/N =1$ \citep{Tremaine2002, Novak2006}. We tested two different implementations of the GLS method: the FORTRAN program LSTSQ \citep{Weiner2006} and the $Python$ program LTS\_LINEFIT \citep{Cappellari2013}. Reassuringly, they return the same best-fit values up to the second decimal digit. We note that LTS\_LINEFIT implements a ``robust'' GLS method that can identify and exclude outliers from the fit, but we do not use this option in our analysis.

The vertical ML method assumes that (1) the intrinsic scatter $\sigma_y$ is Gaussian along the $y$-direction and does not vary along the $x$-direction, (2) $e_x$ and $e_y$ are independent, which is the case for the BTFR. Thus, we can derive the following log-likelihood function \citep[see, e.g.,][]{Pizagno2007}:
\begin{equation}\label{eq:vertical}
 \ln(L) = -\sum_{i}^N \ln(\sqrt{2\pi}\delta_{y, \, i}) - \sum_{i}^{N}\dfrac{\Delta_{y, \, i}^2}{2\delta_{y, \, i}^2},
\end{equation}
where $\Delta_y,\,i$ and $\sigma_{i}$ are the same as in Eq.\,\ref{eq:ydist} and Eq.\,\ref{eq:sigma}, respectively. \citet{Weiner2006} pointed out that the vertical ML method is matematically similar to the GLS method, as one can see by comparing Eq.\,\ref{eq:GLS} and Eq.\,\ref{eq:vertical}. We stress, however, that the GLS method is effectively neglecting the normalization factor of the Gaussian likelihood (the first term in Eq. \ref{eq:vertical}), which contains the free paramter $\sigma_y$. Hence, these two methods are not expected to give the same results. We tested two different implementations of the vertical ML method: the program MLFIT \citep{Weiner2006} and a custom-build program that explores the posterior distributions using $emcee$ \citep{Foreman-Mackey2013}. The two implementations return the same best-fit values, but slightly different errors. This happens because MLFIT does not marginalize over the incertainty in $\sigma_{y}$ (which is not reported) and does a grid search rather than reconstructing the full shape of the joint posterior probability. Our program is superior to MLFIT in this respect.

The orthogonal ML method assumes that (1) the intrinsic scatter $\sigma_\perp$ is Gaussian along the perpendicular direction to the best-fit line, (2) $e_x$ and $e_y$ are independent. The minimum distance between $(x_i, y_i)$ and the best-fit line is
\begin{equation}
\Delta_{\perp,\,i} = \sqrt{\bigg(x_i - \dfrac{x_i + ay_i - ab}{a^2 + 1}\bigg)^2 + \bigg(y_i - b - a\dfrac{x_i+ay_i-ab}{a^2+1}\bigg)^2}.
\end{equation}
The total perpendicular scatter is then given by
\begin{equation}\label{eq:orthogonal}
\delta_{\perp, i} = \sqrt{\bigg(1 - \dfrac{\Delta^2_{\perp, \, i}}{\Delta^2_{y, \,i}}\bigg) e_{xi}^2 +
\dfrac{\Delta_{\perp, \, i}^2}{\Delta^2_{y, \,i}} e_{yi}^2 + \sigma_\perp^2 }.
\end{equation}
Thus, the log-likelihood function can be written as follows:
\begin{equation}
  \ln(L) = -\sum_{i}^N \ln(\sqrt{2\pi}\delta_{\perp, \, i}) - \sum_{i}^{N}\dfrac{\Delta_{\perp, \, i}^2}{2\delta_{\perp, \, i}^2},
\end{equation}
We implemented this method using $emcee$ to explore the posterior distributions. More details are given in Sect.\,\ref{sec:MCMC}.

Figure\,\ref{fig:Fitting} shows the best-fit parameters for the three different methods as a function of the velocity definition. For a given velocity definition, the vertical ML and the GLS methods provide fully consistent values of $\sigma_{y}$ (obviously $\sigma_\perp$ in the orthogonal ML method has a different meaning from $\sigma_{y}$, so the two measurements cannot be compared). The values of $s$ and $I$ are consistent within $\sim$1.5$\sigma$ for the different fitting methods, but there are systematic differences. The vertical ML always provides the shallower relations, while the orthogonal ML always provides the steepest ones. It is clear, however, that the differences between the fitting methods are minor compared to the differences between the velocity definitions. The precise velocity definition plays the major role in setting the properties of the BTFR.
\end{appendix}

\bibliographystyle{mnras}
\bibliography{BTFR2019} 

\bsp	
\label{lastpage}
\end{document}